# Direct correlation between aromatization of carbon-rich organic matter and its visible electronic absorption edge

Nicola Ferralis [a,*], Yun Liu [a], Kyle D. Bake [b], Andrew E. Pomerantz [b], Jeffrey C. Grossman [a]

[a] Department of Materials Science & Engineering, Massachusetts Institute of Technology, Cambridge, Massachusetts, United States

[b] Schlumberger-Doll Research, Cambridge, Massachusetts, United States

The evolution of the electronic absorption edge of type I, II and III kerogen is studied by diffuse reflectance UV-Visible absorption spectroscopy. The functional form of the electronic absorption edge for all kerogens measured is in excellent agreement with the "Urbach tail" phenomenology. The Urbach decay width extracted from the exponential fit within the visible range is strongly correlated with the aliphatic/aromatic ratio in isolated kerogen, regardless of the kerogen type. No correlation is found between the decay width and the average size of aromatic clusters, which is explained in terms of a non-linear increase in optical absorption with increasing size of the aromatic clusters determined by $^{13}$C NMR. Further, absorption spectra calculated with density functional theory calculations on proxy ensemble models of kerogen are in excellent agreement with the experimental results. The correlation of the decay width with conventional maturity indicators such as vitrinite reflectance is found to be good within a particular kerogen type, but not consistent across different kerogen types, reflecting systematic variations in bulk composition for different type kerogen types with the same vitrinite reflectance. Thus, diffuse reflectance visible absorption spectroscopy is presented as a rapid, calibrated and non-destructive method to monitor both the maturity and the chemical composition of kerogen. The chemical insight of kerogen in relation to its optical absorption provided by this methodology may serve for rapid screening of kerogen for electronics and optical devices in place of functionalized produced carbon.

---

[*] Corresponding author. Tel. +1 617-324-0372. E-mail: ferralis@mit.edu (Nicola Ferralis)

## 1. Introduction

Resurgence in interest in the fundamental chemistry of carbon-rich organic matter has been driven by the recent revolution in oil and gas production from organic-rich mudstones, commonly referred to as organic shales [1, 2]. Despite the attention, the molecular chemistry of the insoluble organic matter from these rocks (known as "kerogen") is poorly understood [3]. Kerogen is a carbon-rich organic compound often categorized into three classes. Type I is typically derived from lacustrine sources and may contain the highest hydrogen:carbon (H:C) ratio, Type II is typically derived from marine sources and contains moderate H:C ratio, and Type III is typically derived from terrestrial sources and contains low H:C ratio [4-6]. Beyond the source material, the composition of kerogen is impacted by thermal processes occurring in the subsurface (known as "catagenesis", Fig 1) [7], resulting in a rich and complex framework of carbon-rich organic material, plus fluid hydrocarbons that comprise oil and gas. The extent to which kerogen has been altered by these processes (known as kerogen "thermal maturity", Figure 1) is an important parameter for evaluating shale reservoirs [8, 9].

Common methods of estimating thermal maturity are informative but involve a compromise between accuracy, level of detail, and simplicity [4]. Some methods can quantify maturity but provide no direct characterization of chemical composition (for example, vitrinite reflectance, VRo [10, 11] or Rock-Eval pyrolysis [12, 13]). Chemically specific methods (such as $^{13}$C NMR [14] or elemental analysis) require destructive sample preparation (kerogen isolation through acid demineralization [12, 15]) that is significantly more time consuming than the measurement itself. Furthermore, such alterations to the samples limit the ability to correlate the molecular composition with the local mineralogy.

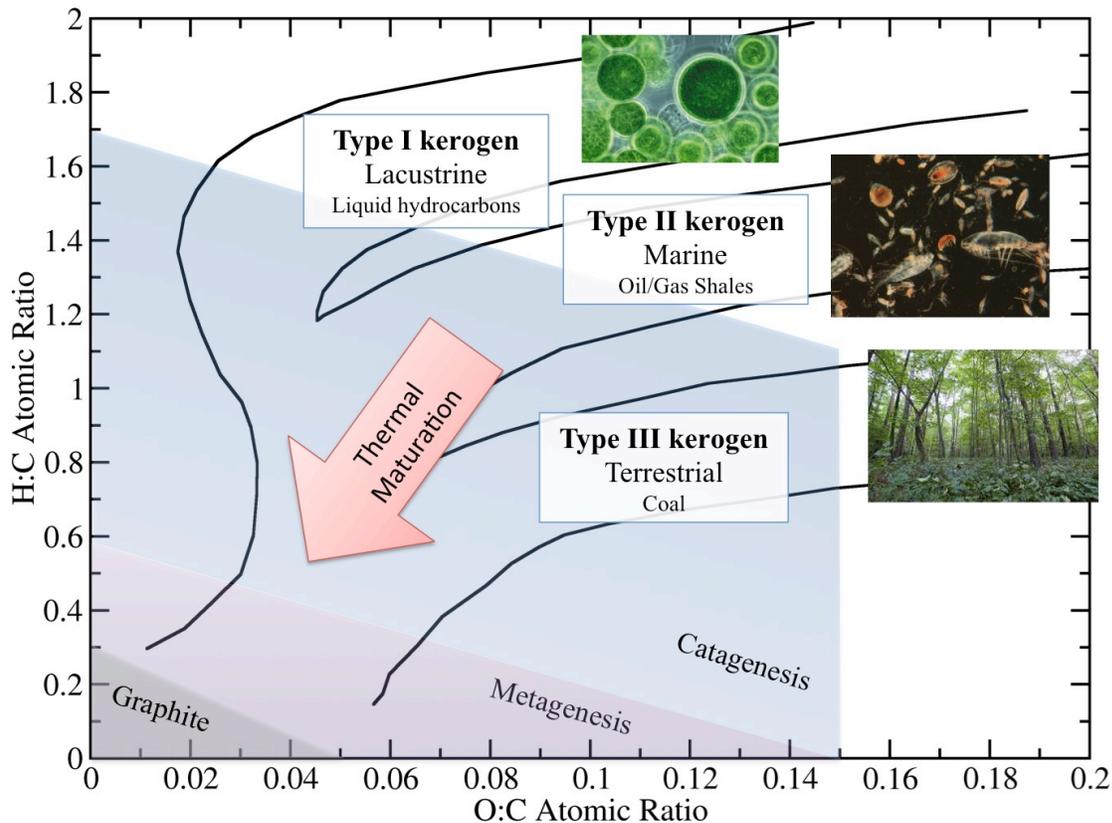

**Figure 1.** Van Krevelen diagram (modified from [6]) highlighting the biological origin of each kerogen type. The shaded area indicates the catagenesis window [6], where biological material is transformed into carbon-rich organic matter under pressure and temperature: thermal maturation.

The importance of understanding the maturity and composition of kerogen motivates new approaches to accurately and non-destructively characterize its detailed chemical structure. Inspired by the role played by the conjugation length and configuration in the optical absorption of aromatic systems, we present a combined experimental and computational approach based on UV-Visible absorption spectroscopy to relate the maturity of kerogen with its conjugation configuration and chemical structure. Benefitting from similar methods that have been used characterize hydrocarbon fluids [16], we employ calibrated techniques that use the Urbach tail phenomenology [17] (described by the exponential dependence of the coefficient of optical absorption with wavelength) to define maturity indicators extracted from kerogen UV-Visible spectra. Through the aid of quantum mechanical simulations for a large set of proxy ensemble models of kerogen, we use these metrics to evaluate and predict with high accuracy the aromatic/aliphatic ratio, through the evolution in concentration of the largest sized aromatic clusters. Finally, we highlight the potential use of UV-Visible diffuse reflectance spectroscopy as a chemically-specific complement to traditional maturity measurements.

**2. Materials and Methods**

2.1 *Experimental*

UV-visible spectra are recorded for type I, II, and III kerogen samples. Type I kerogens were collected from the Green River formation and thermally matured by semi-open pyrolysis, [18, 19] providing estimated VRo [11] between 0.48% and 1.65% (see Supplementary information, Table S1). Type II kerogens were collected from three different formations spanning a range of VRo between 0.55% and 2.2% (see Supplementary Information, Table S2). Type III kerogens were selected from standard coals from the Penn State Coal Bank [20] and Argonne Coal Bank, [21] and their VRo ranged 0.28% to 1.42% (see Supplementary Information, Table S3 and S4). Type I and type II kerogens were isolated from their inorganic matrices by acid demineralization [12]. Coal samples were sufficiently organic rich (at least 70 wt% organic) that they were used without purification. $^{13}$C NMR spectroscopy and elemental analysis were used to derive aromatic and aliphatic content, as shown in Table S2. Spectra of pure graphitic-like materials such as single wall carbon nanotubes and reduced graphene oxide (from Graphene Supermarket) and graphite microplatelets (Fisher Scientific) were recorded to identify the reference baseline for the sensitivity of the optical absorption edge of kerogen to graphitization. The quantification of evolution of the extinction coefficient with the aromatic cluster size was performed using pure aromatic compounds (Sigma-Aldrich): 1-ring (hexamethylbenzene), 2-rings (naphthalene), 3-ring (anthracene), 4-rings (tetracene). UV-visible absorption spectra were acquired using an Agilent Cary 300 UV-Vis spectrometer, with a maximum spectral range between 200 and 800 nm, through a Xe-arc lamp (see Supplemental Information).

2.3 *Proxy Ensemble Models of Kerogen*

While detailed atomic representations of kerogen are reported in the literature [22, 23], new large ensembles of proxy models of kerogen were designed to specifically capture the relation between optical absorption, the substantial degree of statistical variability in conjugation length *and* configuration, and chemical composition of kerogen, as represented in the Van Krevelen diagram (H/C ratio up to 0.6 and O/C ratio up to 0.17, Figure S1). Two approaches were employed to create proxy ensemble models (Figure 2) using Molecular Dynamics (MD) simulations, in order to remove potential initial configuration bias from the preparation of any individual model. The MD simulations in both approaches were carried out using the LAMMPS software package [24]. A reactive force field (ReaxFF) [25] were employed to describe the inter-atomic interactions so that molecules could react freely. In one approach, a given number of C, H and O atoms were randomly mixed in a small simulation cell with periodic boundary conditions, and their bonds further randomized by equilibrating the system at 6000K with NVT ensemble. After heating the system for 5 ns, the structure was cooled down to 500K and re-equilibrated with constant temperature and zero external pressure before the structure was relaxed from *ab initio*. In the other approach, smaller samples were cut from large atomic structures pre-equilibrated from MD simulations. First, 3000 aromatic hydrocarbon molecules with a specific H/C ratio were generated in a simulation cell with at least 4Å separation between each molecule. The total number of atoms in each simulation cell varied from 50,000 to 150,000 depending on the H/C ratio given. The system was then equilibrated under 500 psi pressure and 1500K temperature until the total energy was constant (typically 25 ps of equilibration time). The equilibrated system was then separated into small cubic samples with dimensions 13Å×13Å×13Å each. Oxygen atoms were randomly inserted into each small cubic sample until the desired O/C ratio was reached within that sample. These small samples were then re-equilibrated at 500K within the NVT ensemble to re-establish their periodic boundaries. After the proxy samples were generated from either of the above-mentioned approaches, the structure of each proxy sample was further relaxed with *ab initio* calculations using VASP [26, 27]. During this *ab initio* relaxation step, all the energies were calculated using Density Functional Theory (DFT) with the Perdew-Burke-Ernzerhof [28] (PBE) exchange-correlation functional and a wave function energy cutoff of 500 eV. The positions of the atoms were relaxed until all the forces on the ions were smaller than 0.01 eV/Å.

Absorption spectra were calculated from the Random Phase Approximation (RPA) [29] at the DFT level using VASP. The imaginary part of the frequency dependent dielectric matrix was determined from a summation of the overlap between occupied and empty electronic bands, and 2000 frequency grid points were used to resolve the absorption spectrum of each sample. From the absorption spectra, specific trends (linear and exponential) can be extracted in a manner similar to that employed for the experimental absorption spectra, as described in the next section.

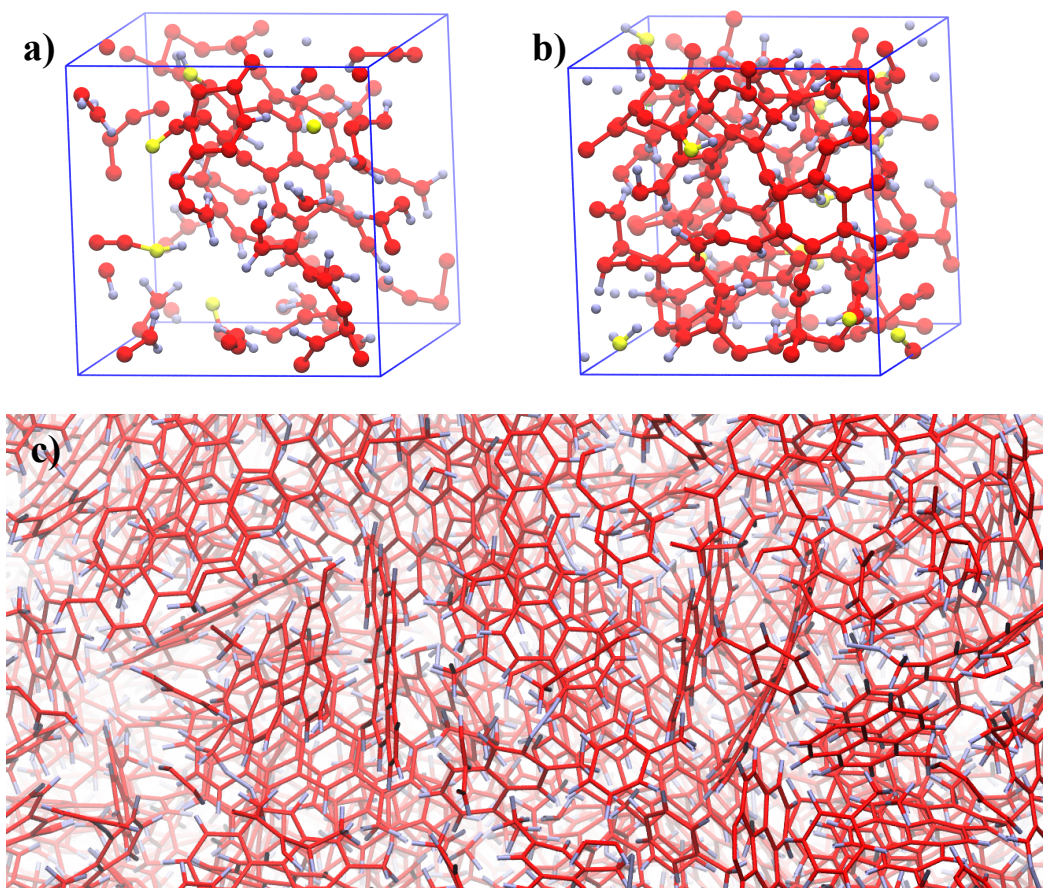

**Figure 2.** Proxy ensemble models of kerogen used in the *ab initio* UV/Vis calculations, generated from two different methods: a) a small piece cut from the large equilibrated system and b) atoms start with random positions in the small box and are then equilibrated. The sample shown in a) has a bulk density of 1.17 g/cm$^3$, H/C ratio of 0.58 and O/C ratio of 0.04. The sample shown in b) has a bulk density of 2.24 g/cm$^3$, H/C ratio of 0.46 and O/C ratio of 0.07. c) A portion of the large equilibrated system which started with a box of random poly aromatic hydrocarbons (PAH) molecules. Grey, red and yellow corresponds to H, C and O atoms, respectively. Bonds are only shown in c) for the ease of viewing. Images are generated using the VMD software package [30].

## 3. Results and Discussion

The widespread use of UV-Visible absorption spectroscopy for characterizing polyaromatic compounds is favored by the highly quantum-mechanical sensitivity of light absorption to chemical conjugation (details in Supporting Information, Chapter S3). For polycyclic aromatic hydrocarbons (PAH), the particle in a box model predicts a reduction of the bandgap as the aromatic cluster size increases. Thus, the spectra of PAHs exhibit sharp edges that shift to the red as the size of the PAH increases (Figure S2). Results for kerogen samples are presented below, first for type II, then type III, then type I. Raw, unprocessed UV-Visible absorption spectra of type II kerogen are shown in Figure 3. The absorption spectra consist of exponential absorption trends with graded slopes in the visible range, a flattening of the absorption in the UV range, and a set of highly reproducible spectral bands around ~3 eV (Figure 3, inset). These bands (such as the Soret band) are typically associated with porphyrins [31] and can be used to assign maceral type (such as liptinite for the Soret band, rather than inertite or vitrinite [32]). Their presence in most cases does not hinder the ability to extract the nearly linear absorption trends in the visible. Absorption trends in the UV range (E > 3 eV) are not determined by conjugation but rather by saturated alkanes [33]. In the interest of relating absorption to aromatic content, we therefore restrict the energy to the visible (1.9 - 3 eV, corresponding to 414 - 652 nm).

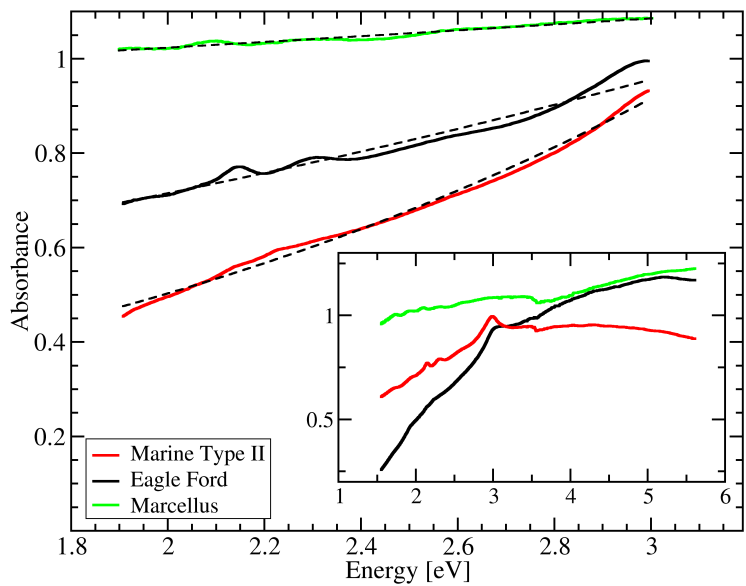

**Figure 3.** Visible absorption spectra of the isolated type II kerogens with exponential fit according to equation 1. (Inset) Full UV-Vis absorption spectra.

The exponential nature of the kerogen spectra contrasts the sharp absorption edges obtained for pure aromatic compounds (Figure S2). Exponential spectra indicate chemically heterogeneous samples and are caused by the overlap of many absorption edges at different wavelengths originating from many diverse light absorbers (chromophores). For example, exponential spectra are observed in crude oils [16] where optical absorption occurs mostly from the asphaltene fraction, which is a complex mixture of single aromatic cores with aliphatic substitutions as described by the Yen-Mullins model [34-36]. Exponential spectra can be described by the Urbach tail phenomenology [16, 37], where the absorption coefficient $\alpha$ is given by:

$$\alpha(\omega) \sim exp\left(\frac{\hbar\omega}{E_0}\right) \qquad (1)$$

According to this description, absorption across the entire range is defined by $E_0$, the Urbach decay width. For semiconductors, $E_0$ represents the thermal excitation of optical absorber sites, which lowers the band gap by exactly the thermal excitation energy ($E_0 = kT$) [16, 38]. However, for crude oils, the decay width $E_0$ corresponds to a measure of the population distribution of chromophores, because larger chromophores have smaller bandgaps while smaller chromophores have larger bandgaps. Visible absorption spectra of isolated type II kerogens are found to conform closely to the Urbach exponential decay (Figure 3), and their decay widths range from 1.8 - 14 eV.

UV-Vis absorption spectra from the type III Argonne Premium Coals are shown in Figure 4a (relevant structural information about the aromatic concentration and configuration [14], are highlighted in Table S4). Because of their high oxygen contents, these samples are in some regards similar to processed amorphous films and graphitic carbonaceous materials, such as polymer-functionalized reduced graphene oxide [39]. As can be seen in the figure, two of the eight coals display highly non-linear behavior: for Pittsburgh No. 8 (HVB) and Pocahontas (LVB), a highly convoluted set of bands characterize the visible portion of the spectra. Similar to the Soret band in the type II kerogens, these bands are not related to the aromatic content in the coals; however, in this case their presence strongly affects the extraction of a meaningful Urbach decay width. Thus, these two coals are not considered in the analysis. This is indeed a general procedure: substantial deviations from the Urbach exponential trends are due to competing absorption processes, and therefore the spectra discarded from the analysis.

The absorption spectra in the visible range for the subset of the Argonne coals are shown in Figure 4b. As with the type II kerogens, these type III kerogens are well described by the Urbach phenomenology. While additional modulation exists in the coal spectra, likely due to residual minerals, the Urbach phenomenology still allows for quantitative analysis.

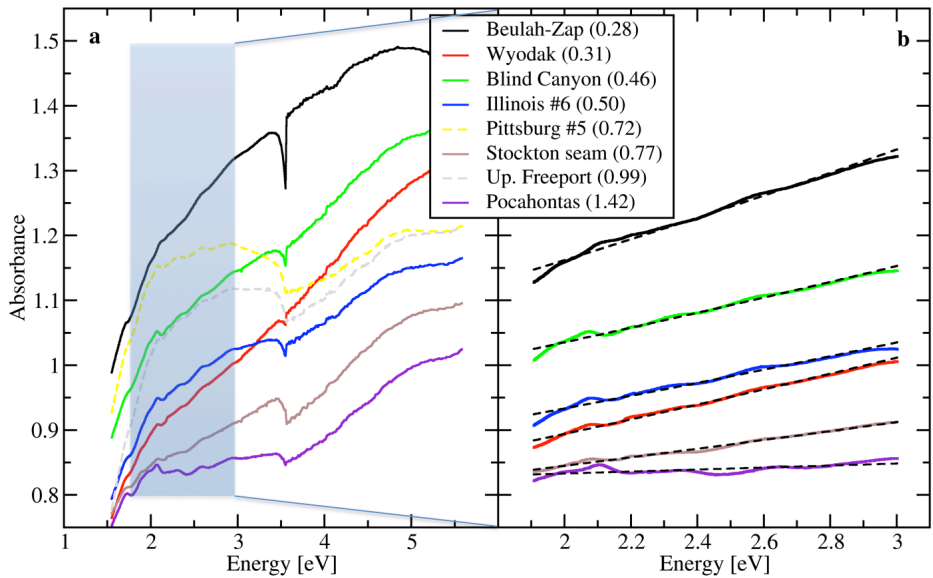

**Figure 4.** a) UV-Visible absorption spectra of selected Argonne Premium Coals. Dashed spectra in Pittsburgh No. 8 (HVB) and Pocahontas (LVB) display a highly non-linear behavior in the visible. The shaded region corresponds to the visible range, plotted in b). b) Visible absorption spectra of the restricted set of Argonne Premium Coal collection. Each spectrum is fitted with an exponential function according to equation 1. For each coal, the corresponding mean vitrinite reflectance value is reported in parentheses.

To understand the chemical significance of $E_0$, fitted values are plotted against average molecular parameters from NMR (Figure 5). $E_0$ is found to correlate tightly with the aliphatic/aromatic ratio for type II and type III kerogens (Figure 5a). Additionally, $E_0$ from proxy ensemble models with different H:C and O:C ratios (and therefore different aromatic contents) correlates strongly with the aliphatic/aromatic ratio (Figure 5b). Even though no fitting parameters are used in the calculations, the trends for measured and simulated samples essentially overlap. Furthermore, the use of two, purposefully different methods in the simulations for the generation of proxy ensemble models for a particular H:C and O:C did not induce a systematic difference in the computed absorption trends. This correlation between $E_o$ and aromaticity, along with known correlations between aromaticity and maturity,[40] form the basis of the use of optical spectroscopy for maturity measurement (discussed below).

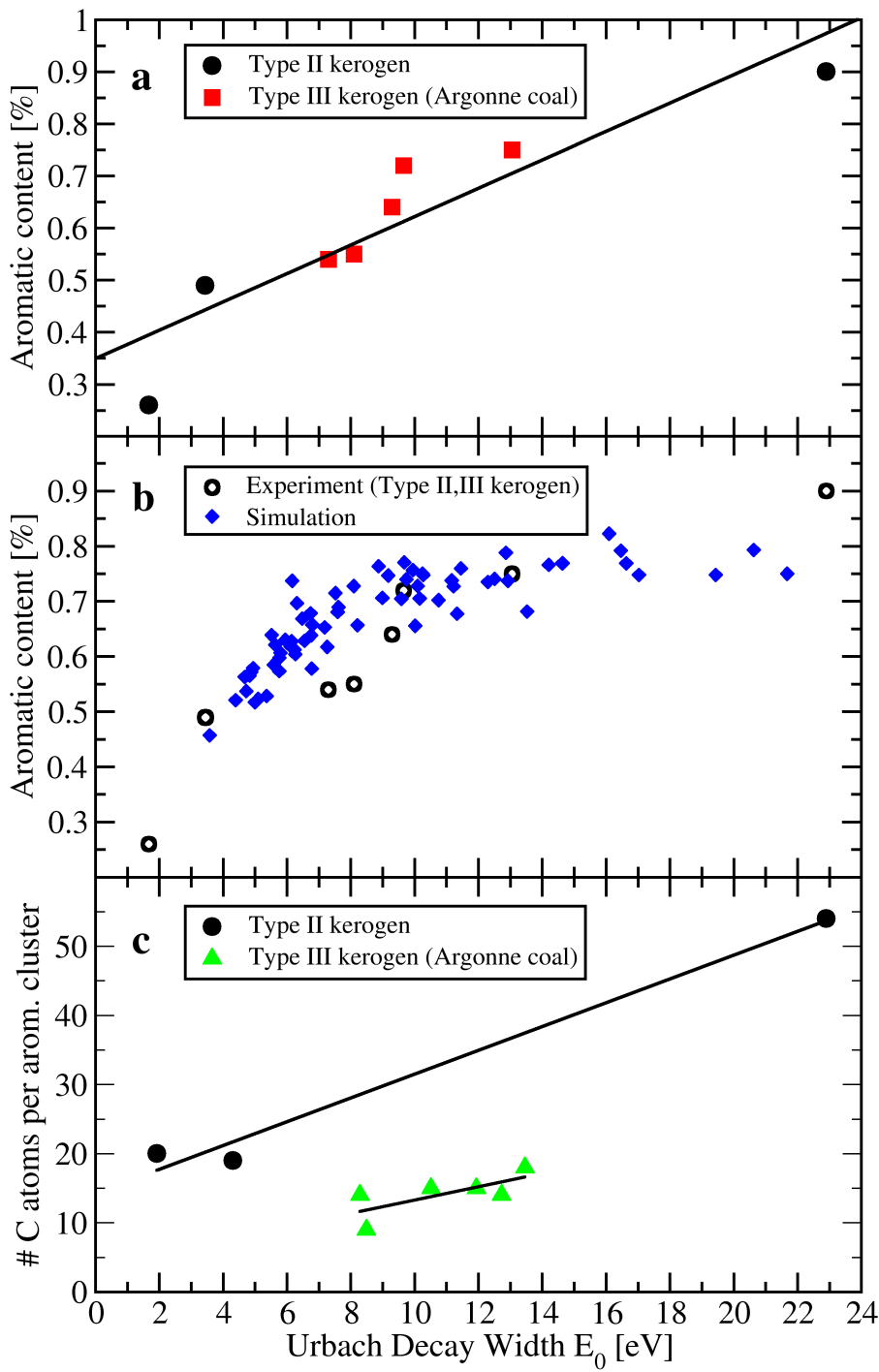

**Figure 5.** a) Correlation between $E_0$ and the aromatic content in type II and type III kerogens (Argonne coals) through a linear fit (Aromatic Content = 0.349 + 0.028*$E_0$, $R^2$=0.917). b) The correlation between $E_0$ and aromatic content is compared between experimentally measured kerogens (type II and III) and calculations of kerogen ensemble proxy models. c) Correlation between $E_0$ and the average aromatic cluster size in type II and type III kerogens.

In crude oils, the correlation between $E_0$ and the aromatic content is further extended to a unique correlation with the distribution of aromatic cluster sizes [16]. In particular, the conjugation length of PAH increases—and therefore the bandgap decreases—with increasing aromatic cluster size (the rate of increase depends on the geometry of the cluster) [14]. If each aromatic cluster contributes equally to the optical absorption (i.e. if the cluster has a similar extinction coefficient regardless of cluster size), a correlation between $E_0$ and the average cluster size (defined as the number of atoms per aromatic cluster size) might be expected. However, a general, type-independent correlation between $E_0$ and the average cluster size is not found for type II and type III kerogens, as shown in Figure 5c. A possible explanation for the lack of correlation is a non-linear increase in the extinction coefficient for increasing aromatic cluster sizes. To verify this hypothesis, Figure 6 presents UV-Vis absorption spectra of mixtures of pure aromatic molecules with well-defined number of aromatic rings (details of the composition of the mixtures in the Supplementary Information). It is found that adding a small amount of a large aromatic cluster to a mixture of smaller cluster results in a marked change in the optical absorption. For example, adding 10% tetracene (4 rings) to a mixture averaging 2 rings (mixture B, Table S5) results in a significant increase in absorption across the full spectra) (mixture C, absorption edge at ~2.2 eV); and adding 6% or 7-10% graphite to mixtures A (average 2.25 rings) and C (average 2.2 rings) respectively nearly doubles the absorption in the pre-edge region, resulting in this mixture with a small average aromatic size having a spectrum that is nearly indistinguishable from that of pure graphite. The non-linear increase in absorption with cluster size might be associated with the larger polarizability per carbon atom in larger aromatic clusters, due to the higher mobility of π electrons [40]. This phenomenon suggest that the absorption spectrum is particularly sensitive to large clusters, such that $E_0$ should be correlated not with the average cluster size but instead to higher moments of the cluster size distribution (which cannot be measured by NMR).

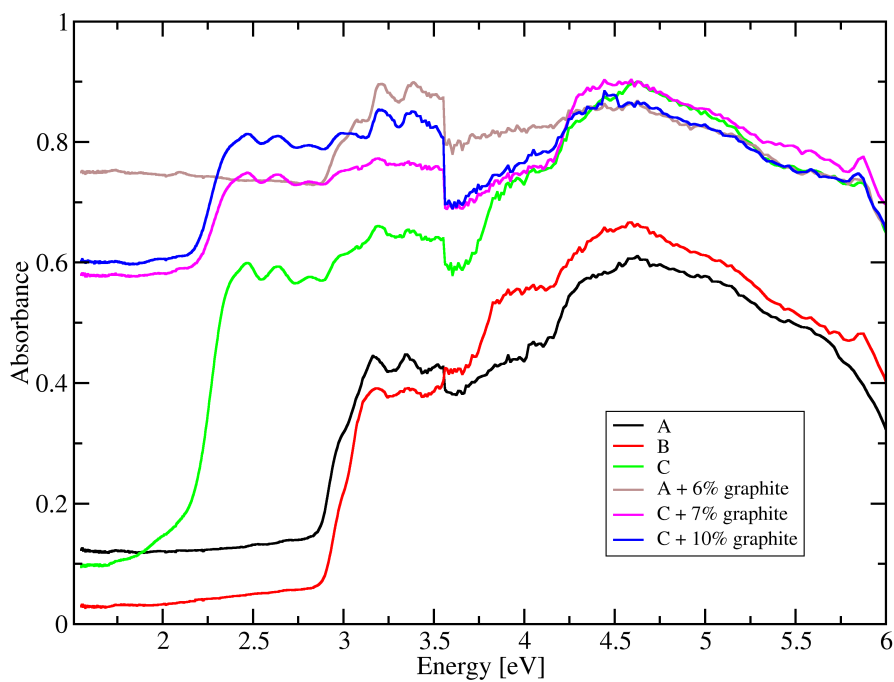

**Figure 6.** UV-Vis absorption spectra from mixtures of pure aromatic compounds and graphite. Mixtures A and B have an average ring size of 3 and 2, respectively. Mixture C is similar to B with added 11% of tetracene (4-ring aromatic). Detailed concentrations of each molecule within each mixture are reported in Table S5.

The use of artificial aromatic mixtures highlights a major difference with kerogen. While in kerogen there is a strong correlation between $E_0$ and aromatic content, such correlation does not necessarily exist for less complex artificial mixtures. For example, different spectra from aromatic mixtures (Figure 6), all of which are nearly 100% aromatic, have different $E_0$. The excellent correlation of $E_0$ with aromatic content in kerogen indicates that kerogen with the largest aromatic clusters also has the highest aromatic content. While the average may change (given the statistical nature of the average over an extensive number of sizes and shapes of aromatic clusters), the tail of the distribution with the largest aromatic clusters may be formed by components that can only be present at a specific maturation stage, regardless of type, resulting also in the correlation with aromatic content. As a result, the level of aromatization of the organic carbonaceous mixture measured through $E_0$ regardless of the kerogen type, provides predictive capabilities of the state of kerogen maturation defined by the degree of aromatization. From a technological standpoint, the largest aromatic clusters in kerogen largely dictate optical absorption, thus providing a crucial parameter for the use of kerogen for light-harvesting applications (i.e., solar cells [41, 42], or photodetectors).

Furthermore, the lack of a unique relation between $E_0$ and the average cluster size suggests that the average size of the aromatic clusters does not represent a consistent indicator of the state of aromatization in kerogen. Kerogen from different biological origins (and therefore of different types) may initially include aromatic compounds with different cluster sizes, which then follow different trajectories during thermal maturation based on their local organic composition. Therefore, when a similar thermal process is applied, a different value for the average size of the aromatic cluster is to be expected. Conversely, from this work we observe that at reservoir conditions, a given kerogen follows a type-specific maturation process that depends on the kerogen original composition. In essence, the predominantly aliphatic nature of biological material before transformation into hydrocarbons (catagenesis) acts as a common denominator across kerogen types, and leads to a common maturation evolution to highly aromatic kerogen. This is consistent with the traditional Van Krevelen diagram, where every kerogen type shows a decrease in H:C ratio (normally associated with an increase in aromaticity) with increasing maturity.

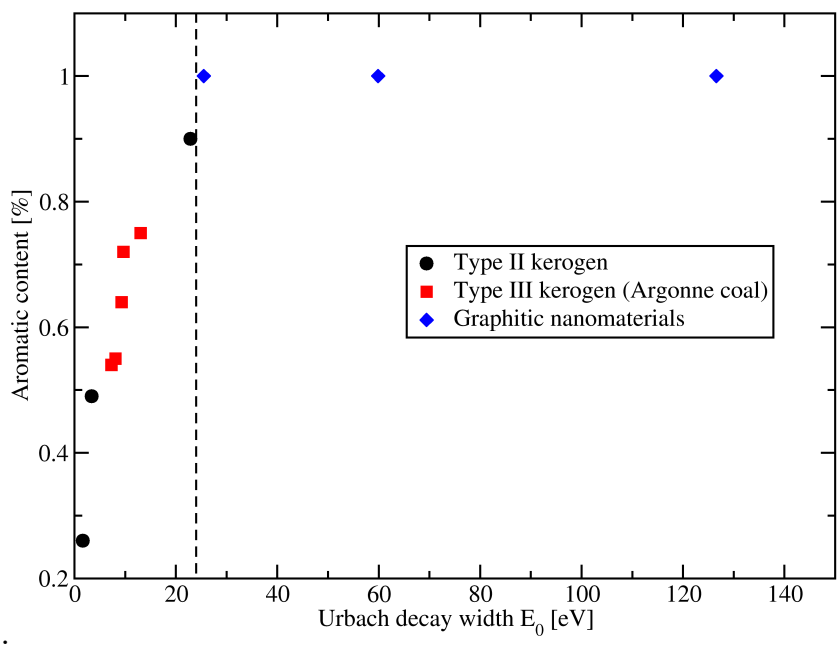

**Figure 7.** The higher limit of the Urbach decay width (dashed vertical line at 25 eV) is provided by the broadband absorption in the visible of graphitic nanomaterials.

As the aromatic concentration in kerogen approaches that of graphitic materials, it is important to evaluate the onset of graphitization. UV-Visible spectra of test graphitic compounds (graphite flakes, reduced graphene oxide, single-wall carbon nanotubes SW-CNT, Figure S4) exhibit extremely shallow slopes of the Urbach tails, asymptotically approaching a horizontal line. The low values for the slope translate into large but unstable values of $E_o$. The linear correlation between $E_o$ and aromaticity holds for kerogen samples with $E_o$ below 25 eV; values of $E_o$ greater than 25 eV indicate nearly fully aromatic, graphitic materials (Figure 7).

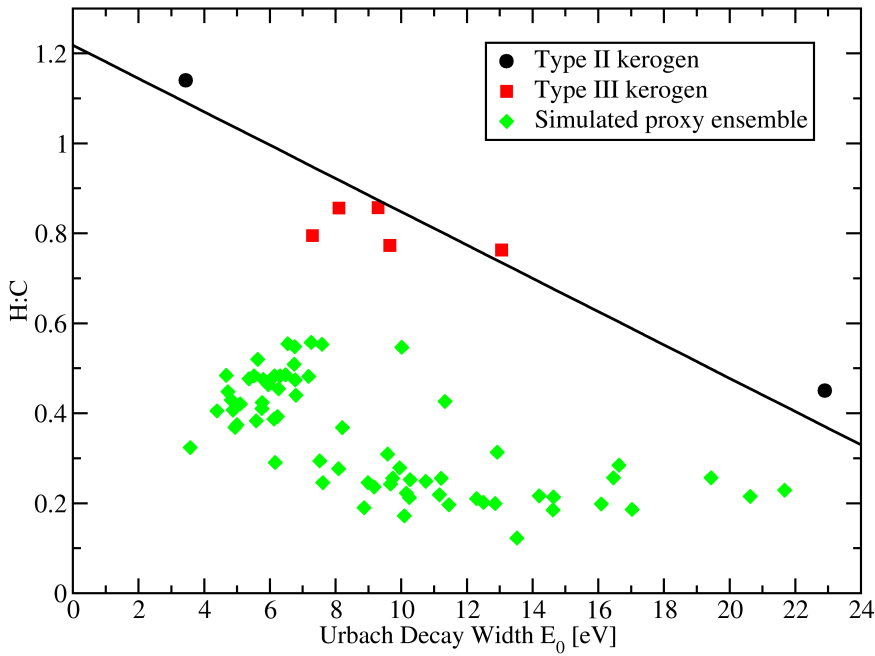

**Figure 8.** Correlation between $E_0$ and H:C ratio is compared between kerogen and simulated kerogen ensemble proxy models. The correlation between the experimentally derived $E_0$ and H:C ratio in type II and type III kerogens is found to be linear (black line, H:C = 1.218 + 0.037*$E_0$, $R^2$=0.92).

In many carbonaceous materials, aromaticity is strongly correlated to the H:C ratio, because aromatic systems typically contain lower H:C ratios than aliphatic systems [20]. Consistently, a strong correlation between $E_o$ and H:C ratio is observed for type II and type III kerogen (Figure 8a). When the experimental correlation is compared to that obtained from kerogen ensemble models (Figure 8b), the H:C ratio in the latter appears to be underestimated, despite a similar trend. We attribute the lower H:C in the proxy ensemble models to subtle differences in the structure and functional groups (such as the lack of methyl groups) compared to kerogen. Kerogen can contain abundant methyl groups formed by breaking weak beta bonds in alkyl-aromatic structures, and that kinetic effect is not included in the model construction [43]. Methyl groups do not contribute significantly to the optical absorption in the visible range (which is dominated by the aromatic clusters) but their presence increases the H:C ratio. The relative lack of methyl groups in the models therefore decreases the H:C ratio compared to kerogens with the same $E_0$.

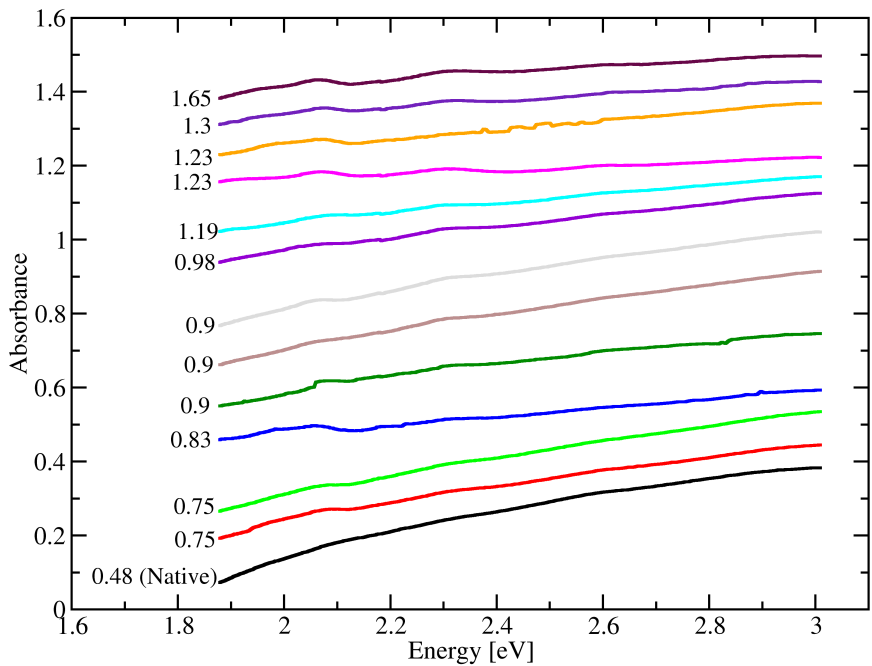

**Figure 9.** Visible absorption spectra of artificially matured Type I kerogens (Green River Formation). Values on the left of each spectra correspond to the estimated reflectance value (%Ro).

Maturation typically results in a decrease in H:C ratio and an increase in aromaticity [40]. The tight correlations of $E_0$ with H:C ratio and with aromaticity suggests that UV-visible spectroscopy might be used as a rapid measurement of maturity. Maturity is most commonly measured by vitrinite reflectance (VRo) [13], in which the amount of light reflectance of a particular organic carbonaceous maceral, vitrinite, is used (when present in the natural carbonaceous material) to indicate maturity. When vitrinite is not present in kerogen, kinetically derived models have been developed to estimate the effective reflectance values (such as the "Easy%Ro" method [11]), based on thermal history. Figure 9 shows the UV-Visible spectra of artificially matured type I kerogens. The use of artificially matured type I kerogen, starting from the same initial kerogen, is similar to the use of processed carbon to understand its thermally induced aromatization, as it tests the consistency of estimates in optical response based on the chemical kinetics during thermal processing or maturation [17]. The spectra follow the Urbach tail phenomenology, consistent with the other kerogen types.

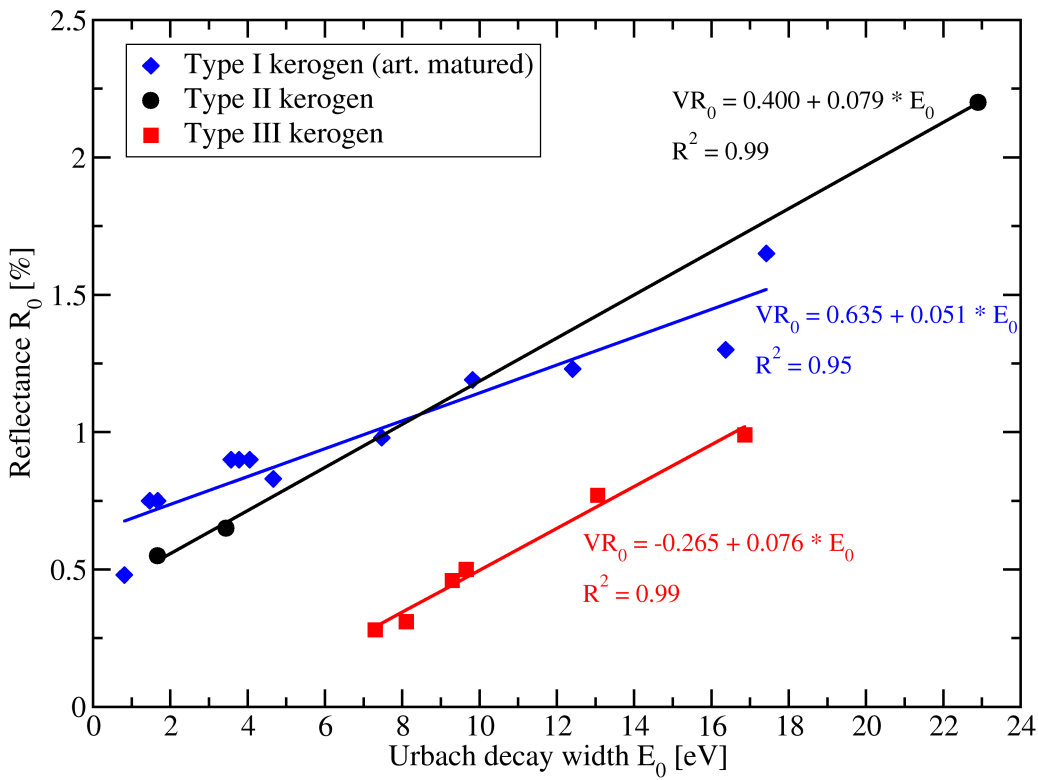

**Figure 10.** Correlation between the Urbach decay width, $E_0$ and reflectance ($VR_0$ for type II, III; $EasyR_o$ [11] for type I kerogen).

Figure 10 compares $E_0$ with the measured or simulated reflectance values for types I, II, and III kerogen. Tight, linear correlations are found for each kerogen type, spanning the entire maturity range investigated. However, each type is found to have a unique correlation between $E_0$ and $VRo$. This result contrasts the correlation with aromaticity, where a single correlation between $E_0$ and aromatic content was found to hold for multiple kerogen types (Figure 5).

All kerogen types become more aromatic with increasing maturation [44]. However, different kerogen types have different rates of aromatization and different distributions of aromatic compounds at a given maturity [16]. These variations are expected based on the different geochemical compositions of the initial organic compounds. By extension, it is to be expected that the rate of aromatization of processed carbon (such as graphene oxide [45]) depends on the initial covalently bonded oxygen content. While the Urbach decay width captures the chemical composition of the entire sample, vitrinite reflectance provides a measurement of the thermal evolution of only the vitrinite maceral. While the restriction to a single maceral enables comparison of VRo values across different kerogen types, it results in different kerogen types having different bulk compositions even at the same VRo. Because $E_0$ captures the composition of the entire kerogen, as opposed to a single maceral, different correlations with VRo are obtained for different kerogen types. At high maturity (VRo > 2), all kerogen types transform into similar, highly graphitic materials [6], and consistently the correlations of $E_0$ with VRo trend towards convergence.

## 4. Conclusions

Unlocking the potential of carbon rich organic matter in organic shales for hydrocarbon production as well as for their applications in novel nanostructure carbon-based devices will benefit from a deeper understanding of its complex chemical and structural composition. Yet, the chemical heterogeneity of kerogen and its evolution upon thermal maturation impose severe limits in the application of conventional methods for characterization. Current methods either rely on subsets of the mixture (enabling comparison across kerogen types but lacking direct connection with chemical composition) or on extensive material preparation (prohibiting routine application). Here, diffuse reflectance visible absorption spectroscopy is presented as a rapid, calibrated and non-destructive method to monitor the chemical evolution of kerogen through the process of natural and artificial maturation. We show, both experimentally and computationally, a strong correlation between the visible optical absorption with the aliphatic/aromatic ratio, justified within the Urbach phenomenology. We underscore that the universality of such correlation (compared to the non-unique relation between VRo and kerogen composition) highlights the intimate connection between the chemical evolution of the largest aromatic clusters and the evolution of the aromatic/aliphatic ratio achieved through thermal maturation. The theoretical framework presented in this work indicates the key role played by the statistical nature of the evolution of different organic natural carbon materials during thermal maturation, as captured by the Urbach tail phenomenology. The Urbach decay width can serve as a universal indicator for the state of aromatization and, when calibrated for kerogen type, for thermal maturation in carbon-rich organic matter.

The observed non-linear increase in optical absorption with the aromatic cluster size is essential to capture the state of thermal maturation and also provides new technological insight for the design of light gathering devices based on natural and processed carbon. Together with the Urbach decay width, it serves as a validation step in the design of micro-structurally and chemically accurate kerogen models, which can act as theoretical bases for the use of kerogen as a light-gathering nanoscale electronic material. Understanding the structure of kerogen is crucial beyond the hydrocarbon production standpoint. Inspired by the use of produced nanostructured carbon materials in electronics [46, 47] and solar power generation [41, 42], the chemical, physical and structural diversity of natural carbon from kerogens and its potential tunability through sample selection, offers a set of yet unexplored and unconventional opportunities for its use as active materials for nano and micro-scale electronic devices.

## 5. Acknowledgements

We gratefully acknowledge Shell Oil Company for providing kerogen type II samples and AMSO for providing kerogen type I samples. We wish to thank the X-Shale consortium and its members Schlumberger and Shell Oil Company for partial funding of this effort under the MIT-Energy Initiative. We are grateful to NERSC for providing computing resources for this work.